\documentclass[aps,prl,showpacs,twocolumn,superscriptaddress]{revtex4-2}%

\usepackage{graphicx}
\usepackage{dcolumn}
\usepackage{bm}

\usepackage{float}  

\usepackage{graphicx}
\usepackage{graphics}
\usepackage[caption=false]{subfig}
\usepackage{color}
\usepackage[english]{babel}
\usepackage[utf8]{inputenc}
\usepackage{ae}
\usepackage{latexsym}
\usepackage{dcolumn}
\usepackage{bm}
\usepackage[table]{xcolor}
\usepackage{array}
\usepackage{url}
\usepackage{epsf}
\usepackage{epsfig}
\usepackage{pstricks,pst-grad}
\usepackage[pdftex,colorlinks]{hyperref}
\usepackage{amsmath}
\usepackage{amssymb}
\usepackage{ragged2e}
\usepackage{comment}
\usepackage{adjustbox}
\usepackage[normalem]{ulem}
\usepackage{capt-of}

\begin{document}

\title{
Microphase Separation and Ionic Aggregation in Oppositely Charged End-Functionalized Polymer Blends
}

\author{Chao Duan}
\affiliation{Division of Chemistry and Chemical Engineering, California Institute of Technology, Pasadena, California 91125, United States}

\author{Zhen-Gang Wang}
\email {zgw@caltech.edu}\affiliation{Division of Chemistry and Chemical Engineering, California Institute of Technology, Pasadena, California 91125, United States}

\date{\today}

\begin{abstract}

Using an ion-correlation-augmented self-consistent field theory (IC-SCFT), we study the phase behavior of binary blends of A and B polymers carrying single end groups of opposite charge.
Under the low dielectric conditions typical for organic polymer melts, strong ion correlations drive association of multiple chain ends, leading to emergent multichain architectures that go beyond a simple ion-paired diblock copolymer picture previously used to interpret recent experiments. At moderate to strong polymer segregation, the resulting phase diagram differs considerably from this effective-diblock prediction: ion clustering broadens the lamellar and cylindrical windows while suppressing the network phases, the reverse of the trend for end-charged diblock copolymers. 
At weak segregation, the two polymers remain miscible even as ion clustering persists, and decreasing chain length drives the clusters themselves through a sphere-to-cylinder-to-lamella shape transition, connecting continuously to the aggregation behavior of classical terminally charged ionomers in the limit when one of the polymer species shrinks to just the charged end. These results reveal that ion clustering — beyond simple pairwise bonding — shapes the self-assembly of end-charged polymer blends.

\end{abstract}

\maketitle

\section{Introduction}

Polymer blends underpin diverse technologies, ranging from plastics recycling, solid-state electrolytes, to functional materials 
\cite{Garcia_2017,Sullivan_2022,Caradant_2021,Blatt_2021,Cooper_2023,Clarke_2023,Zhang_2025}.
A central challenge for most chemically dissimilar commodity polymers is their intrinsic immiscibility and low entropy of mixing, leading to macrophase separation with poor mechanical and optical properties \cite{Eagan_2017,Kuwashiro_2026} and thereby limiting their practical reuse in sustainable applications \cite{Vozniak_2025,Zhu_2016,Mohanty_2022}.
Introducing only a small number of oppositely charged groups can effectively suppress macrophase separation and compatibilize otherwise immiscible polymers through strong ionic association~\cite{Russell_2002,Pispas_2003,Huh_2006,Zhang_2014,Fredrickson_2022,Le_2023,Das_2025,Qin_2026,Xu_2025,Edmund_2025}.






Recently, Xie et al. studied blends of polystyrene (PS) and polydimethylsiloxane (PDMS) in which each chain carries a single functional end group, with proton transfer between the end groups of the two polymers rendering the PS and PDMS chain ends oppositely charged \cite{XieS_2023}.
The resulting ionic association between 
terminal charges was interpreted as forming an effective diblock copolymer.
In the strong-bonding regime, where proton transfer occurs for most chains, well-ordered lamellar microphases were observed.
Notably, blends with longer chains tend to form less-ordered microstructures, indicating a reduced effective segregation strength, in contrast to neutral block copolymers \cite{XieS_2023}.
For compositionally asymmetric PS/PDMS blends, 
they further observed hexagonally packed cylindrical morphologies, and a cylinder-to-lamellae transition upon strengthening the ionic association \cite{Karnaukh_2024}.
These observations imply that ionic association influences the phase behavior beyond simply linking unlike polymers into effective diblocks. 

Polymer field-theoretic approaches constitute an efficient theoretical framework for elucidating the behavior of blends of oppositely charged polymers, wherein ionic association is represented either via phenomenological binding descriptions or through an explicit treatment of electrostatic interactions. 
Pair-binding models, which are widely employed for telechelic and supramolecular polymer systems~\cite{Vigil_2023,Danielsen_2023,Balzer_2025,Balzer_2026}, represent oppositely charged moieties by a reversible bound–unbound equilibrium governed by an effective binding free energy.
Although such models reproduce several experimental trends, including ionic-bond-controlled domain swelling, microphase-to-macrophase transitions, and chain-length-dependent compatibilization~\cite{Fredrickson_2022,XieS_2023}, they do not capture the collective ion clustering induced by strong ion correlations in the intrinsically low-dielectric environment of polymer melts~\cite{Middleton_2017,Shen_2021}. 
Alternatively, electrostatic interactions have been treated explicitly within the polymer field-theoretical framework~\cite{Brereton_1990,Rumyantsev_2019,Rumyantsev_2020,Grzetic_2021,Le_2022}.
Existing approaches successfully capture electrostatically driven compatibilization and microphase separation in oppositely charged polymer blends. However, they are based primarily on mean-field, Born solvation energy, or weak-fluctuation descriptions of the electrostatics, leaving the strong ion-correlation effects, particularly ion clustering and the resulting self-assembly, essentially unresolved.

In this letter, we apply our ion-correlation-augmented self-consistent field theory (IC-SCFT) \cite{Duan2025PEBrush,Duan:2026aa} to study the phase behavior of binary blends of A and B homopolymers in which each A chain carries a single cationic end group and each B chain a single anionic end group, with no added counterions.
Our approach incorporates the variational Gaussian renormalized fluctuation theory for electrostatics \cite{Wang:2010wk} into SCFT for polymers \cite{Fredrickson_Book2006}.
We find that strong ion correlations drive the association of multiple charged termini, leading to emergent multichain architectures. Strikingly, this ion clustering reshapes the phase behavior, producing both qualitatively altered polymer microphase separation and a distinct regime of ionic aggregation within otherwise disordered polymers.
These phenomena cannot be captured by a simple ion-paired diblock copolymer picture.



\section{Model and Methodology}

We consider an incompressible melt of polymer blend in a volume of $V$, with equal numbers of A and B chains.
Each A- and B-polymer is functionalized at one chain end by an ion with valency $z_A=+1$ and $z_B=-1$, respectively, while the other end remains neutral.
For simplicity, we assume that all polymers have an equal Kuhn length $b$ and segment volume $v$. The number of segments on each $K$-polymer is specified as $N_K$ ($K=A,B$).
The first segment of each chain is designated as the ionic group.
The Flory--Huggins parameter describing the immiscibility between A- and B-segment is $\chi$.
We also assume identical dielectric constants for both polymers, specified as $\varepsilon_r$.
All charges are described by a finite-sized ion model given by the distribution function $h_K({\bf r},{\bf r}^{\prime})$ with the Born radius $a_K$ 
\cite{Wang:2010wk,Agrawal:2022ux,Duan2025PEBrush}.
We further assume an equal Born radius of end-ions for both polymers, $a_+=a_-=a$.
Here, we take $h_K$ to be Gaussian for convenience. We adopt the discrete Gaussian chain model such that each charged segment is explicitly considered as an ion with finite-spread charge.

Following the standard field theoretical approach for polymers \cite{Fredrickson_Book2006} and the nonperturbative variational approach to account for the electrostatic fluctuation by introducing a Gaussian reference action \cite{Wang:2010wk,Agrawal:2022ux,Duan2025PEBrush},
a set of self-consistent equations can be obtained for volume fraction $\phi_K(\bf r)$ and the conjugate fields $\omega_K(\bf r)$ of the $K$-polymers ($K=A,B$), mean electric potential $\psi(\bf r)$, ion concentration $c_K(\bf r)$ and self energy $u_K(\bf r)$ of charges, 
and electrostatic correlation function $G(\bf r,{\bf r}^{\prime})$. We refer interested readers to our earlier paper for the detailed derivation \cite{Duan2025PEBrush}. 
\begin{subequations}
\begin{align}
\phi_K({\bf r}) &= \frac{\overline{\phi}_K}{N_K Q_K} \sum_{i=1}^{N_K} q_K^{\dagger}({\bf r};i) e^{V_K({\bf r};i)} q_K({\bf r};i) 
\end{align}
\begin{align}
\omega_A({\bf r}) -\omega_B({\bf r})  & = \chi [\phi_B({\bf r})-\phi_A({\bf r})]
\end{align}
\begin{align}\label{PB}
-\epsilon \nabla^2\psi({\bf r}) &= \sum_{K=A,B} z_K c_K({\bf r})
\end{align}
\begin{align}
c_K({\bf r}) &= \frac{\overline{\phi}_K}{N_K vQ_K} q_K^{\dagger}({\bf r};1) e^{V_K({\bf r};1)} q_K({\bf r};1)
\end{align}
\begin{align}\label{uK}
u_K({\bf r}) &= \frac{z^2_K}{2} \int d{\bf r^{\prime}} d{\bf r^{\prime\prime}} h_K({\bf r},{\bf r}^{\prime}) G({\bf r}^{\prime},{\bf r}^{\prime\prime}) h_K({\bf r^{\prime\prime}},{\bf r})
\end{align}
\begin{align}\label{GreenFunc}
-\epsilon \nabla^2_{\bf r} G({\bf r},{\bf r}^{\prime}) + 2I({\bf r}) G({\bf r},{\bf r}^{\prime}) &= \delta({\bf r}-{\bf r}^{\prime})
\end{align}
\end{subequations}
where $\epsilon =k_BT \varepsilon_0 \varepsilon_r /e^2$ is the scaled permittivity with $\varepsilon_0$ the vacuum permittivity, $e$ the elementary charge.
$I({\bf r})=(1/2)\sum_{K=A,B} z^2_K c_K({\bf r})$ is the local ionic strength. 
$\overline{\phi}_K$ is the average volume fraction of the $K$-polymers ($K= A, B$).
Note that even with a uniform dielectric permittivity, the presence of ionic aggregates formed by terminal charges results in inhomogeneous ion correlation contributions, including in the nominally “disordered” phases.
$Q_K=V^{-1} \int d{\bf r}~ q^{\dagger}_K({\bf r};1)$ is the corresponding single-chain partition function, while $q_K({\bf r};i)$ and $q_K^{\dagger}({\bf r};i)$ are the forward and backward chain propagators, respectively \cite{Fredrickson_Book2006}, determined by 
\begin{subequations}\label{q_K}
\begin{align} 
q_K({\bf r};i) &= e^{-V_K({\bf r};i)} \int d{\bf r^{\prime}} \Phi ({\bf r} - {\bf r^{\prime}}) q_K({\bf r^{\prime}};i-1) \\
q_K^{\dagger}({\bf r};i) &= e^{-V_K({\bf r};i)} \int d{\bf r^{\prime}} \Phi ({\bf r} - {\bf r^{\prime}}) q_K^{\dagger}({\bf r^{\prime}};i+1)
\end{align}
\end{subequations}
with the initial conditions $q_K({\bf r};1) = e^{-V_K({\bf r};1)}$ and $q_K^{\dagger}({\bf r};N_K)=e^{-V_K({\bf r};N_K)}$.
$V_K({\bf r};i)$ is the total interaction field experienced by the $i$-th segment of a $K$-polymer.
$V_K({\bf r};i)=\omega_K({\bf r})+z_K\psi({\bf r})+u_K({\bf r})$ for the charged segment $i=1$ and $V_K({\bf r};i)=\omega_K({\bf r})$ otherwise.
$\Phi ({\bf r} - {\bf r^{\prime}})$ in Eq. \ref{q_K} is the Gaussian bond transition probability \cite{Fredrickson_Book2006}.

The resulting canonical free energy density of the system $f=v F/(Vk_BT)$ is given by
\begin{align}\label{FreeE}
&f = - \frac{\overline{\phi}_A}{N_A} \ln Q_A
-  \frac{\overline{\phi}_B}{N_B} \ln Q_B
\nonumber\\
&+ \frac{1}{V} \int d{\bf r} \left ( \chi \phi_A\phi_B -\omega_A \phi_A -\omega_B \phi_B + \frac{v \epsilon}{2}\psi\nabla^2\psi  \right) \nonumber\\
&+ \frac{v}{V} \int d{\bf r} \sum_{K=A, B} c_K({\bf r}) \left[ \int^1_0 d{\tau} u_K({\bf r};\tau)-u_K({\bf r}) \right]
\end{align}
The last term in Eq. \ref{FreeE} is the ion correlation energy derived from the ``charging method'' with $\tau$ ($0 \le \tau \le 1$) being the ``charging parameter''. \cite{Wang_2015,Agrawal:2022ux} The intermediate self energy $u_K({\bf r};\tau)$ during the ``charging process'' is calculated from the corresponding Green function $G({\bf r},{\bf r}^{\prime};\tau)$, which is solved by Eq. \ref{GreenFunc} with $I({\bf r})$ replaced by $\tau I({\bf r})$.


\section{Results and Discussion}

\begin{figure*}[t]
\centering
\includegraphics[width=0.98\textwidth]{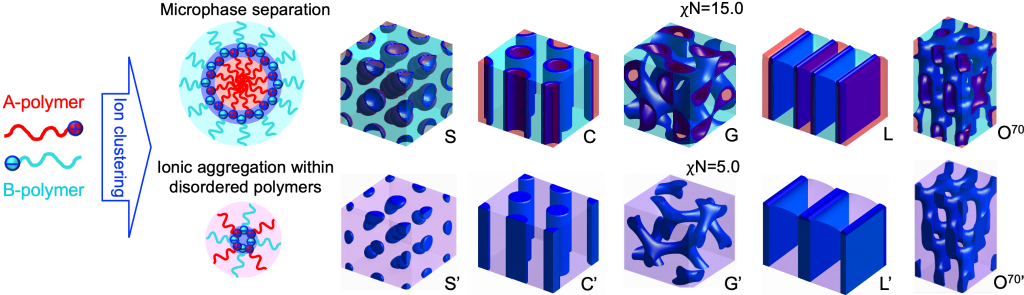}
\caption{Schematic of ion clustering in a binary blend of end-ion-functionalized polymers, together with isosurface plots of polymer and charge densities for the candidate structures: microphase-separated body-centered cubic spheres (S), hexagonal cylinders (C), double-gyroid networks (G), lamellae (L), and orthorhombic {\it Fddd} networks (O$^{70}$) at $\chi N=15.0$; and corresponding plots for the ionic aggregates within disordered polymers at $\chi N=5.0$ (S$^{\prime}$, C$^{\prime}$, G$^{\prime}$, L$^{\prime}$, and O$^{70 \prime}$).
$N=100$ and $a=0.5$nm. 
}
\label{fig1}
\end{figure*}


Ion correlations drive the clustering of end ions and effectively link multiple A- and B-polymers at their termini, leading to an emergent neutral 
multichain architecture and various self-assembled structures (Fig.~\ref{fig1}).
The candidate morphologies include body-centered-cubic spheres (S), hexagonally packed cylinders (C), double-gyroid networks (G), lamellae (L), and orthorhombic {\it Fddd} networks (O$^{70}$) (Fig.~\ref{fig1}).
These structures span the principal morphologies relevant to the microphase behavior in the comformationally symmetric neutral block copolymers. 
Other structures, such as face-centered-cubic spheres, Frank--Kasper phases, and more complex networks~\cite{Dorfman_2021,Xie_2023,Xu_2024,You2026}, are not considered in this letter.
For convenient comparison with neutral diblock copolymers, 
we define a ``total chain length'' as $N=N_A+N_B$ for a pair of A- and B-polymers, with the A-polymer volume fraction given by $f_A=N_A/N$. 
Throughout this letter, the Kuhn length and segment volume are set as $b = 1.0$nm and $v = 0.2$ nm$^3$, respectively.
The Bjerrum length is taken to be a constant $l_B=22.4$nm, corresponding to $\epsilon_r=2.5$ at room temperature, representative of a low-dielectric polymer melt.
The coupling parameter characterizing the Coulomb strength can be estimated as $\Gamma= l_B/a \approx 45$ for ions with radius $a=0.5$nm, indicating that the system lies in the strong ion-correlation regime~\cite{Levin_2002}.

\begin{figure}[h!]
\centering
\includegraphics[width=0.48\textwidth]{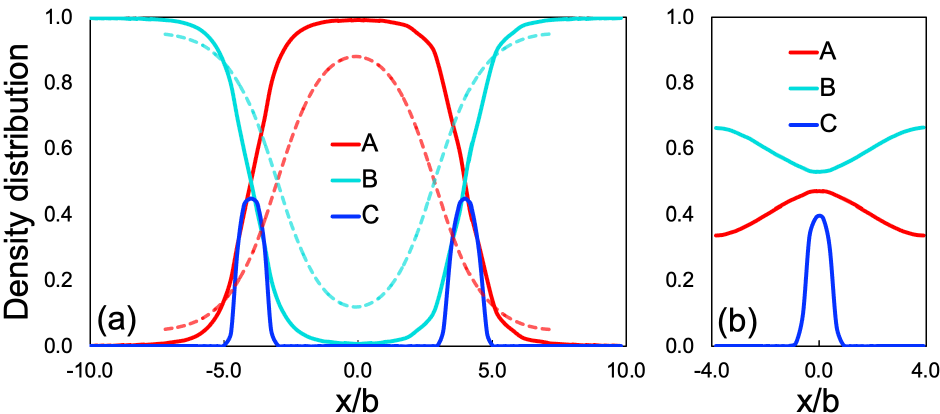}
\caption{1D density distribution (volume-fraction) of A-, B-polymers, and terminal charge C of either A- or B-polymer over one lamellar period, with the distance expressed in units of the Kuhn length $b$: (a) L phase at $\chi N=15.0$, and (b) L$^{\prime}$ phase at $\chi N=5.0$. $N=100$, $f_A=0.4$. The dashed lines indicate corresponding density profiles for neutral diblock copolymers under otherwise identical parameters. Note that the polymer density includes the terminal segment.
}
\label{fig2}
\end{figure}

The polymer-density and end-ion profiles in Fig.~\ref{fig1} exhibit two qualitatively distinct scenarios of self-assembly.
At stronger polymer segregation strength ($\chi N=15.0$), the immiscible A- and B-polymers form their respective microdomains, with the end ions localized at the A/B interfaces. 
Ionic association therefore accompanies the polymeric microphase separation and 
is constrained by the interfacial geometry.
The 1D density profiles further show that the width of interfacial ion-layer is approximately 1.0nm (Fig.~\ref{fig2}a), comparable to the chosen ion size ($a=0.5$nm), consistent with tightly localized ion clustering schematically illustrated in Fig.~\ref{fig1}.
The A/B interface is much sharper than that of the corresponding neutral diblock copolymer (Fig.~\ref{fig2}a), indicating a substantially enhanced effective segregation strength between two polymers. 
This is because strong ion correlation favors dense ionic clustering at the A/B interface, leading to crowded segmental packing and increased chain stretching near the charged termini, which effectively sharpens the A/B interface. In addition, the strong cohesive energy due to ion correlation resists areal expansion, further increasing the effective interfacial tension between the A and B domains.
A comparison with the neutral A$_m$B$_m$ miktoarm star copolymers shows that this interface is as sharp as that of A$_{20}$B$_{20}$ (Fig. S1).
Although the present A- and B-polymers do not necessarily form exact star-like copolymers (e.g., bottlebrush- or platebrush-like copolymers), these structural features strongly support the picture that ion clustering connects multiple polymers into emergent multichain 
architectures rather than simple ion-paired diblocks ($m=1$) .

At weaker segregation ($\chi N=5.0$), in contrast, the A- and B-polymers remain mixed, whereas the end ions remain strongly correlated and organize into aggregates with various morphologies (Fig.~\ref{fig1}).
Compared with the stronger segregation regime, the end ions are distributed around the centroid of the structure; the markedly reduced domain size provides further evidence that A- and B-polymers are much less stretched and stay essentially disordered
(Fig.~\ref{fig2}).
The persistence of ionic aggregation without polymeric microphase separation demonstrates that ionic association can independently drive spatial organization, beyond the simple picture of ion-paired block copolymer.





\begin{figure}[h!]
\centering
\includegraphics[width=0.48\textwidth]{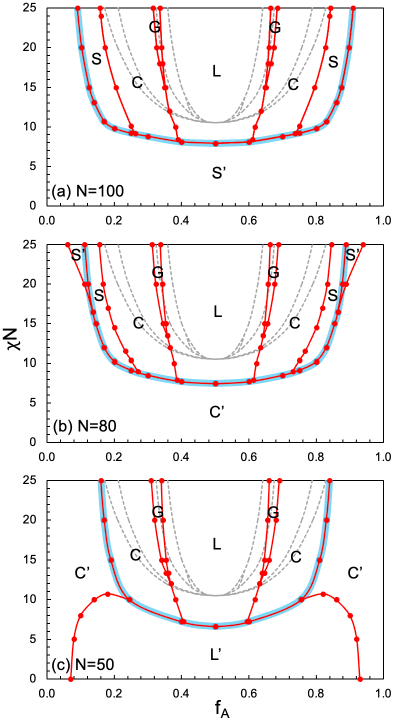}
\caption{Phase diagrams in the $f_{\rm A}$--$\chi N$ plane for the candidate structures listed in Fig.~\ref{fig1}: (a) $N=100$, (b) $N=80$, and (c) $N=50$.
The boundaries between regimes of ordered and disordered polymers are 
highlighted in bold blue.
Phase diagram of the neutral linear diblock copolymers is shown by the gray dashed curves.
}
\label{fig3}
\end{figure}

The phase behavior also differs markedly from that of neutral diblock copolymers.
The relative stability of candidate
structures is determined by comparing their equilibrium free energies (Fig.~\ref{figS2}),
from which full phase diagrams in the $f_A$--$\chi N$ plane can be constructed.
Across various chain lengths, 
two distinct regimes can be identified: an upper polymer-ordered regime, including S, C, G, and L; and a lower polymer-disordered regime including S$^{\prime}$, C$^{\prime}$, and L$^{\prime}$ (Fig.~\ref{fig3}).
In the upper regime, the overall shape resembles that of neutral diblock copolymers, but the phase boundaries are substantially altered.
The polymeric order-disorder transition (ODT) at $f_A=0.5$ is found at a lower segregation strength than that of neutral diblock copolymers, $(\chi N)_{\rm ODT}<(\chi N)_{\rm nc}=10.5$ (Fig.~\ref{fig3}).
This can be understood by noting that ion-correlation-induced multichain association effectively increases the total chain length, thereby strengthening polymeric segregation.
As chain length increases from $N=$ 50 to 80 and 100, corresponding $(\chi N)_{\rm ODT}$ shifts from 6.6 to 7.5 and 7.9, respectively.
This trend implies that longer chains are more difficult to microphase separate, in contrast to the neutral block copolymer picture but in good agreement with the experiments reported by the Segalman group~\cite{XieS_2023}.
Direct transitions from disordered to ordered structures can also be noted, such as ${\rm S}^{\prime} \rightarrow $~C and ${\rm S}^{\prime} \rightarrow $~L without passing through S and C, respectively (Fig.~\ref{fig3}a), distinct from the mean-field phase behavior of neutral block copolymers.

Besides, ordered structures exhibit 
unexpected phase selections in this regime.
The stability windows of L and C are broadened, whereas the G-network phase is suppressed and O$^{70}$-network phase completely disappears (Fig.~\ref{fig3}).
These candidate network structures become unstable, or at most metastable, over a significant portion of the weak-to-intermediate segregation strengths, in which they would normally occupy stable regions between the C and L phases in neutral copolymers.
As chain length increases from $N=$ 50 to 100, the onset of stable G phase shifts from $\chi N=12.2$ to 15.0, while its stability window narrows, 
indicating disfavored formation of G-network in longer chains.
In contrast, the S phase exhibits an opposite $N$-dependence (Fig.~\ref{fig3}): at $N=$ 100, its stability region is considerably widened relative to neutral diblocks; as $N$ decreases, the S-window gradually narrows and vanishes at sufficiently short chain length ($N=50$).
These phenomena are not simple displacements of conventional block-copolymer phase boundaries, but a systematic restructuring due to ion-correlation effects.


Another qualitatively different behavior appears in the lower $\chi N$ regime, where end-ions arrange into aggregates within disordered polymers (Fig.~\ref{fig3}).
It is clear that their structures change systematically with chain length.
For longer chains ($N=100$), ions form isolated spheres. As $N$ decreases, they transform first into cylindrical and then into lamellar structures, producing a ``${\rm S}^{\prime} \rightarrow {\rm C}^{\prime} \rightarrow {\rm L}^{\prime}$'' sequence.
This ionic aggregation is analogous to amphiphilic self-assembly, with the end ion playing the role of a ``hydrophobic head'' and the neutral polymer chain acting as a ``hydrophilic tail''.
The average density of end-ions increases with decreasing chain length, favoring a smaller spontaneous curvature toward the ionic termini, therefore, otherwise discrete ionic clusters tend to connect and form progressively more extended aggregates~\cite{Nyrkova_1993,Semenov_1995}.
%
For the S$^{\prime}$ phase, the aggregation number of polymer chains in each ionic cluster is insensitive to the Flory--Huggins parameter $\chi$ (Fig.~\ref{figS3}a), but decreases appreciably with increasing chain length $N$ (Fig.~\ref{figS3}b), reflecting the increased entropic cost for forming larger aggregates with increasing chain lengths.
Furthermore, neither G$^{\prime}$ nor O$^{70 \prime}$ network can be stable throughout the entire phase diagram.

This polymer-disordered regime, marked by the formation of ion aggregates without microphase order, lies entirely outside the conventional block-copolymer framework and instead establishes a direct connection to ionomer systems~\cite{Middleton_2017,Shen_2021}. In the limit of extreme compositional asymmetry, $N_A = 1$ or $N_B = 1$, one component effectively reduces to a small, mobile ion, leaving a polymer chain bearing an opposite charge at its terminus. The resulting system is thus a terminally charged ionomer; similar ionic aggregate morphologies have indeed been reported in ionomers with other charge distributions (e.g., randomly placed along the backbone) by both experiments and molecular dynamics (MD) simulations~\cite{Seitz_2010,Hall_2011,Hall_2012,Bolintineanu_2013,Sampath_2018,Balzer_2022}. Our IC-SCFT approach thus captures both polymer microphase separation and ionomer-like ionic aggregation within a single, unified framework, revealing two distinct regimes of ion-correlation-driven self-assembly in oppositely end-charged polymer blends. Within this polymer-disordered regime, our mean-field calculations predict ordered, essentially periodic arrangements of the ion aggregates. In practice, however, fluctuations in the shape (including local curvature) and position of the ion aggregates in the low elastic modulus environment of the surrounding disordered polymer matrix, would likely destroy this long-range periodicity, yielding instead the liquid-like, short-range-ordered arrangement of ion aggregates characteristic of real ionomers.

Several key features of the phase behavior can be elucidated by analyzing the ion-correlation energy.
In our IC-SCFT, the free energy density can be decomposed into three contributions: the mixing enthalpy contribution $\Delta u$, which accounts for the interfacial energy; the chain stretching energy $-T\Delta s$; and the 
ion-correlation energy $\Delta f_e$, each measured relative to the completely homogeneous state~\cite{Duan:2026aa}.
In the polymer-ordered regime, $\Delta u$ decreases monotonically while $-T\Delta s$ increases monotonically, along the sequence of progressively lower-curvature morphologies, ``S $\rightarrow$ C $\rightarrow$ G $\rightarrow$ L'' ({Figs.~\ref{figS4}a,~\ref{figS4}b), consistent with the behavior of neutral block copolymers~\cite{Matsen_1996b,Matsen_1997}.
In contrast, $\Delta f_e$ exhibits a nontrivial dependence on morphology.
In particular, near the C--L boundary, we find $\Delta f_e$(G) $>\Delta f_e$(C) $>\Delta f_e$(L) (Fig.~\ref{figS4}c), rationalizing the disappearance of stable G region at weaker segregation strengths in Fig.~\ref{fig3}.
At larger segregation strengths, however, $\Delta u$ and $-T\Delta s$ grow in magnitude, whereas $\Delta f_e$ varies more weakly and becomes comparatively less important (Fig.~\ref{figS4}d), allowing G to gradually recover its stability.
Near the S--C boundary, we likewise find $\Delta f_e$(S) $<\Delta f_e$(C) (Fig.~\ref{figS5}c), accounting for the broadened stable S-region at $N=$ 100 and 80 (Figs.~\ref{fig3}a,~\ref{fig3}b).
%

\begin{figure}[h]
\centering
\includegraphics[width=0.48\textwidth]{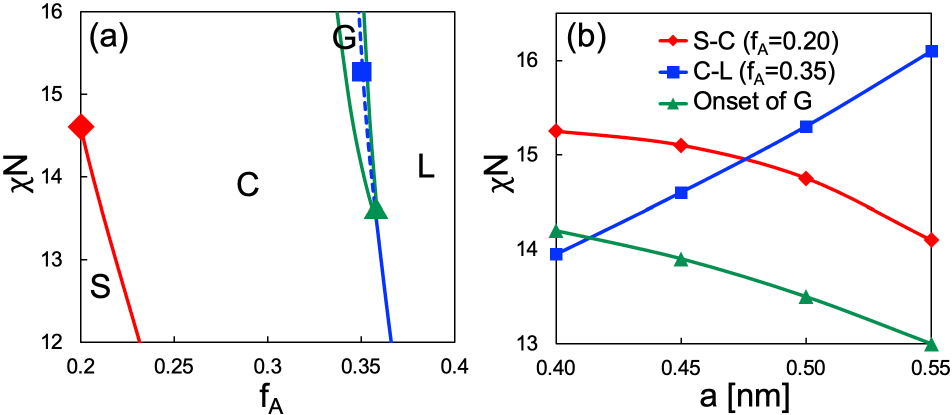}
\caption{(a) Zoomed-in phase diagram corresponding to Fig.~\ref{fig3}b for $N=80$ and $a=0.5$nm, showing the phase boundaries among S, C, G and L. The C--L boundary within the stable region of G phase is indicated by a dashed line.
The S--C and C--L transition points at $f_A=$ 0.20 and 0.35, respectively, together with the onset $\chi N$ of the G phase, are marked by symbols. (b) $\chi N$-shifts of the three marked characteristic points with ion radius $a$.
}
\label{fig4}
\end{figure}

The role of ion correlations in regulating the relative stability of the ordered phases can be further examined by varying the ion size.
Smaller ions, characterized by a smaller Born radius $a$, experience stronger ion correlations.
Figure~\ref{fig4}b summarizes the resulting shifts of three characteristic phase transition points marked in Fig.~\ref{fig4}a.
As $a$ decreases from 0.55 to 0.40nm, the onset of stable G phase shifts to higher $\chi N$, indicating more pronounced suppression of G with stronger ion correlations. 
The C--L transition shifts to lower $\chi N$ with decreasing $a$, consistent with the lower ion-correlation energy of L over C, i.e., $\Delta f_e$(L) $<\Delta f_e$(C).
This trend is also in agreement with the experimentally observed ``C $\rightarrow$ L'' transition under stronger ionic bonding \cite{Karnaukh_2024}.
The opposite shift of the S--C transition reflects the corresponding stabilization of S relative to C, consistent with $\Delta f_e$(S) $<\Delta f_e$(C).
Together, these 
ion-size dependent phase transitions
consistently demonstrate that stronger ion correlations shift the phase boundaries toward ordered structures with lower ion-correlation free energies.
The ion size also affects the order–disorder transition. In particular, decreasing the ion radius increases the transition point $(\chi N)_{\rm ODT}$ between the L and S$^{\prime}$ phases at $f_A=0.5$ (Fig.~\ref{figS6}), because stronger ion correlations preferentially stabilize the more compact ionic aggregates in the S$^{\prime}$ phase.
Based on the same consideration, we can anticipate that smaller ion size favors the association of ions into more extended aggregates, thereby promoting transition from S$^{\prime}$ to C$^{\prime}$ and eventually to L$^{\prime}$. Similar trends are expected with decreasing dielectric permittivity.

Finally, it is instructive to consider the phase behavior in the limit of long polymer chains. In the polymer-disordered regime, as $N$ increases, the decreased concentration of charged termini increasingly favors simple ion pairing over multichain aggregation due to the greater loss of translational entropy associated with the latter (which scales logarithmically with $N$); this trend is consistent with reduced aggregation number with increasing $N$ (Fig.~\ref{figS3}b). In this large-$N$ limit, the association behavior can be described by a simple pairing equilibrium~\cite{Fredrickson_2022,XieS_2023}.  In the polymer-ordered regime, however, ionic clustering localized in the A/B interfacial region still persists, which results in increased chain stretching normal to the A/B interface, contributes to additional effective surface tension, and possibly incurs bending cost of the interface. Consequently, although simple ion pairing is expected to become the dominant mechanism in the polymer-disordered phase in the large-$N$ limit, the energetics of the polymer-ordered morphologies will still influence their relative stability as well as their stability with respect to the polymer-disordered phase.

\section{Conclusions}

In summary, using an ion-correlation-augmented self-consistent field theory (IC-SCFT), we have shown that strong ion correlations qualitatively reshape the phase behavior of binary polymer blends with oppositely charged termini.
Strong ion correlations drive the association of multiple end ions into emergent multichain architectures, producing two distinct regimes of self-assembly.
In the polymer-ordered regime, ion clustering broadens the lamellar and cylindrical stability windows while suppressing network phases.
We demonstrate that the relative stability can be consistently regulated by morphology-dependent ion-correlation energies.
In the polymer-disordered regime, polymers remain mixed, whereas the end ions retain strong correlations and form ordered aggregates that evolve from spherical to cylindrical and lamellar structures with decreasing chain length.
This regime also connects continuously to the behavior of ionomer systems in the compositional asymmetry limit.
Our results show that the collective ion clustering, rather than simple pairwise bonding alone, provides the pivotal mechanism for the self-assembly of oppositely end-charged polymer blends.

Beyond the present study, our IC-SCFT framework can be extended to investigate polymers with different architectures, compositions, charge patterns, and dielectric inhomogeneities.
In experimental systems involving an acid-base equilibrium, incomplete proton transfer leaves a fraction of polymers uncharged~\cite{Russell_2002,Pispas_2003,Huh_2006,Zhang_2014,Fredrickson_2022,Xu_2025,Edmund_2025,XieS_2023,Karnaukh_2024}, potentially giving rise to transitions between microphase- and macrophase-separated states, as well as Lifshitz-point and microemulsion-like structures~\cite{Bates_1997,Duchs_2003,Shim_2019b}.
Furthermore, additional mobile ions, due to screening effect, can lead to similar behavior.
For sufficiently long chains, higher levels of organization, including intradomain ionic segregation, may emerge. 
These subjects form fertile grounds for further investigations.


\section{Supporting Information}
The Supporting Information includes comparison of 1D density profile with miktoarm star copolymers;
equilibrium free energy curves; dependence of aggregation number on the Flory--Huggins parameter and chain length; decomposition of free energy into individual contributions; and effect of ion size on the order-disorder transition point..

\section{ACKNOWLEDGEMENTS}
This research is supported by funding from the Hong Kong Quantum AI Lab and AIR@InnoHK of the Hong Kong Government, and from the Caltech Marcus Center for Theoretical Chemistry.

\bibliography{Tex_Refs}

\clearpage

\setcounter{figure}{0}
\renewcommand{\thefigure}{S\arabic{figure}}

\section*{Supporting Information}

\begin{figure}[H]
\centering
\includegraphics[width=0.48\textwidth]{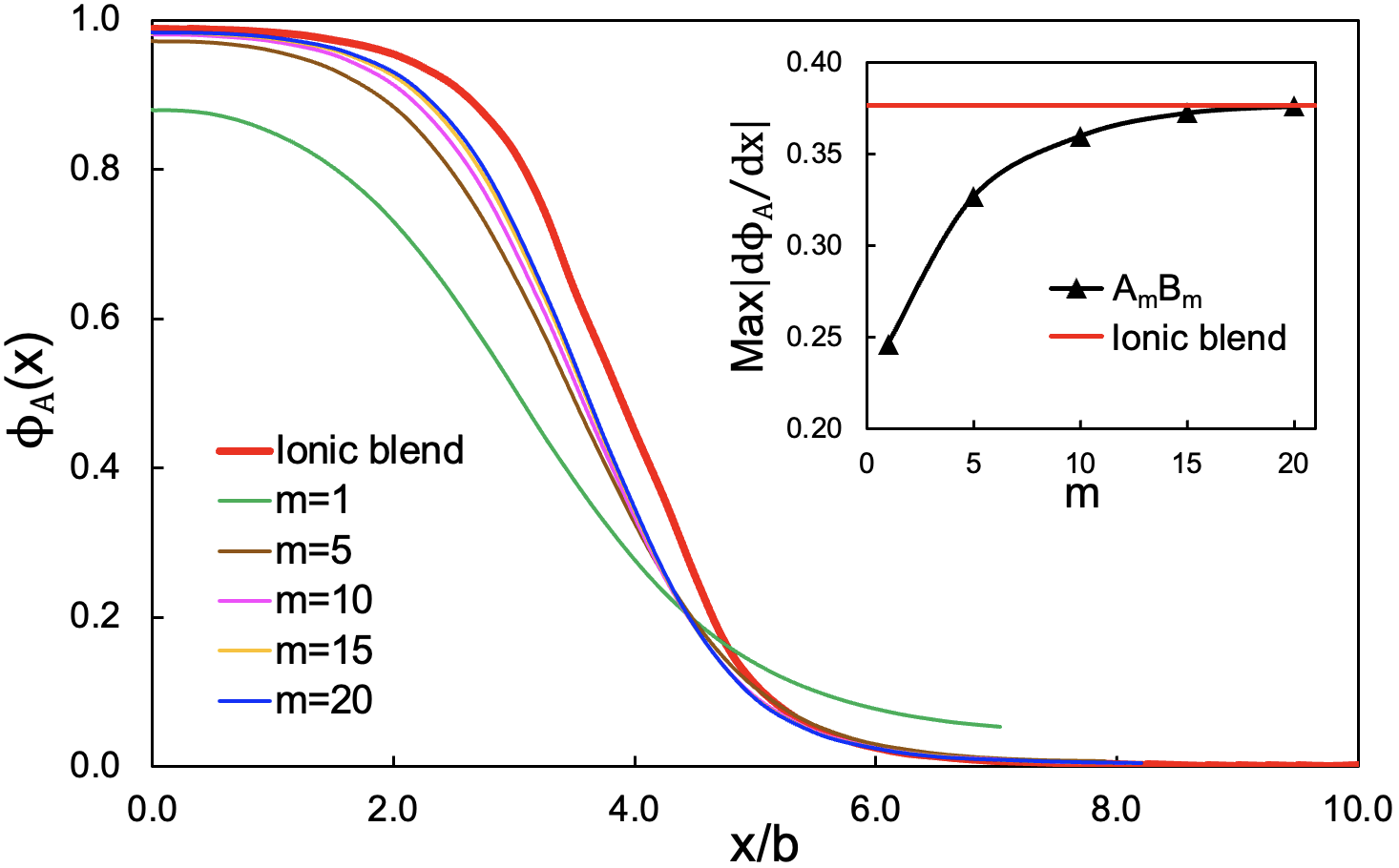}
\caption{Comparison of the 1D density profiles of the A component $\phi_A$ between end-ionic polymer blends and neutral A$_m$B$_m$ miktoarm star copolymers over half of the equilibrium lamellar domain. The inset shows the interfacial sharpness, measured by the maximum slope of $\phi_A$, as a function of $m$.
$\chi N=15.0$, $N=100$, and $f_A=0.4$.
For A$_m$B$_m$ copolymers, $N$ denotes the total segment number of one pair of A- and B-arms.
}
\label{figS1}
\end{figure}

\begin{figure}[H]
\centering
\includegraphics[width=0.42\textwidth]{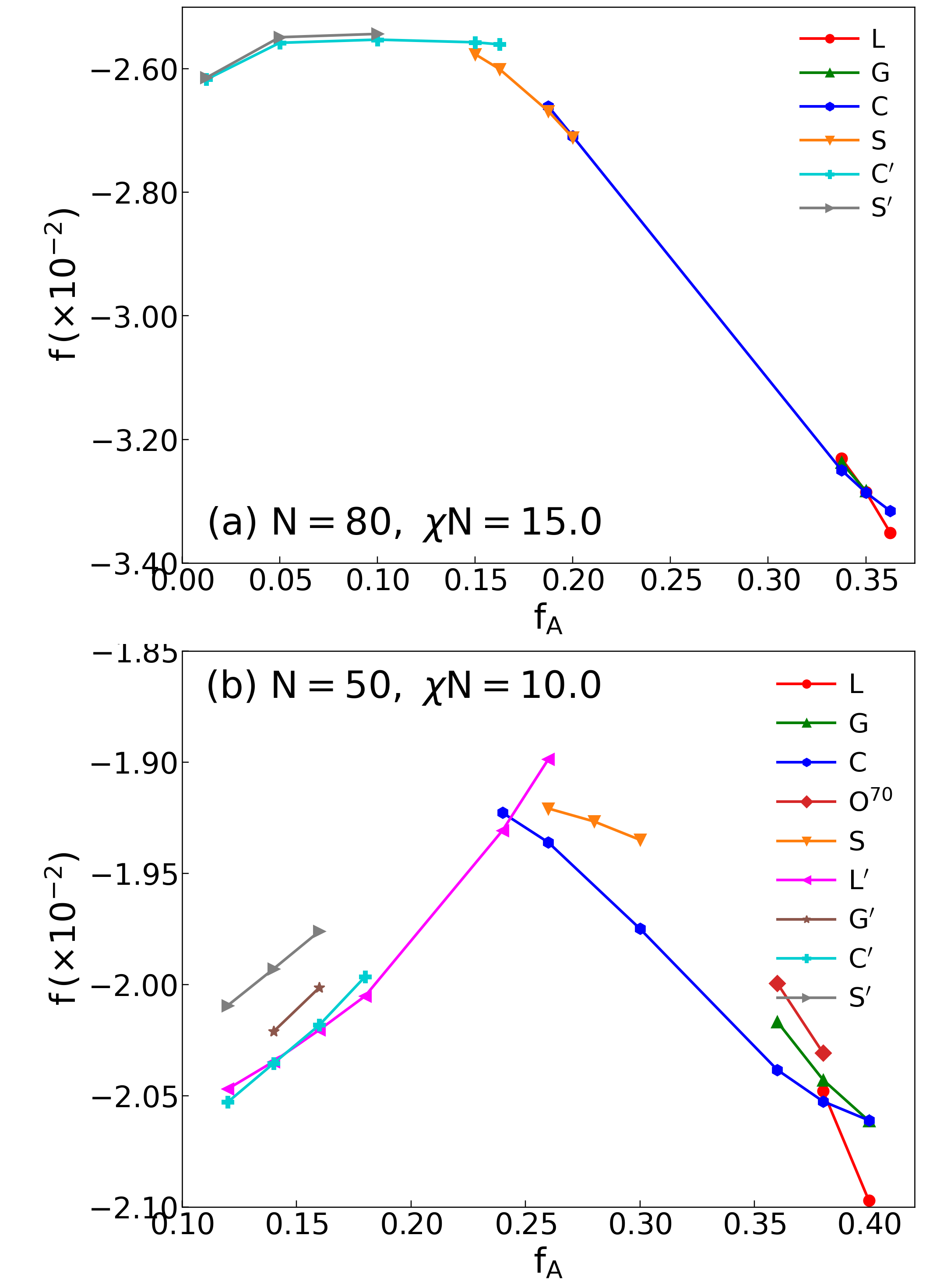}
\caption{Free energy density $f$ (Eq.~\ref{fig3} in the main text) as a function of A-fraction $f_A$ for the candidate structures at (a) $N=80$ and $\chi N=15.0$, and (b) $N=50$ and $\chi N=10.0$.
}
\label{figS2}
\end{figure}

\begin{figure}[H]
\centering
\includegraphics[width=0.48\textwidth]{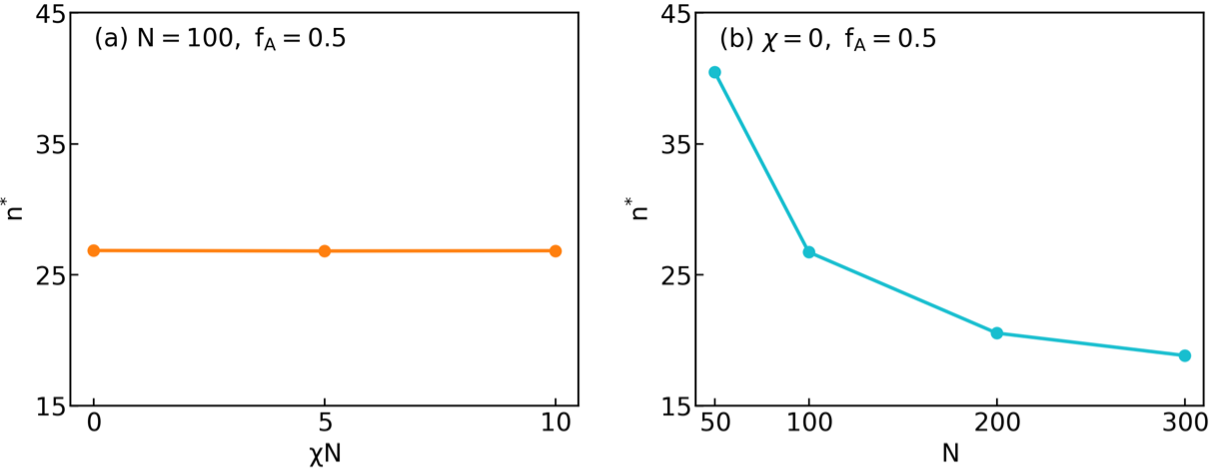}
\caption{Dependence of the aggregation number $n^*$ in the S$^{\prime}$ phase on (a) Flory--Huggins interaction parameter $\chi N$ at fixed $N=100$, and (b) chain length $N$ at fixed $\chi=0$. $f_A=0.5$.
}
\label{figS3}
\end{figure}

\begin{figure}[H]
\centering
\includegraphics[width=0.48\textwidth]{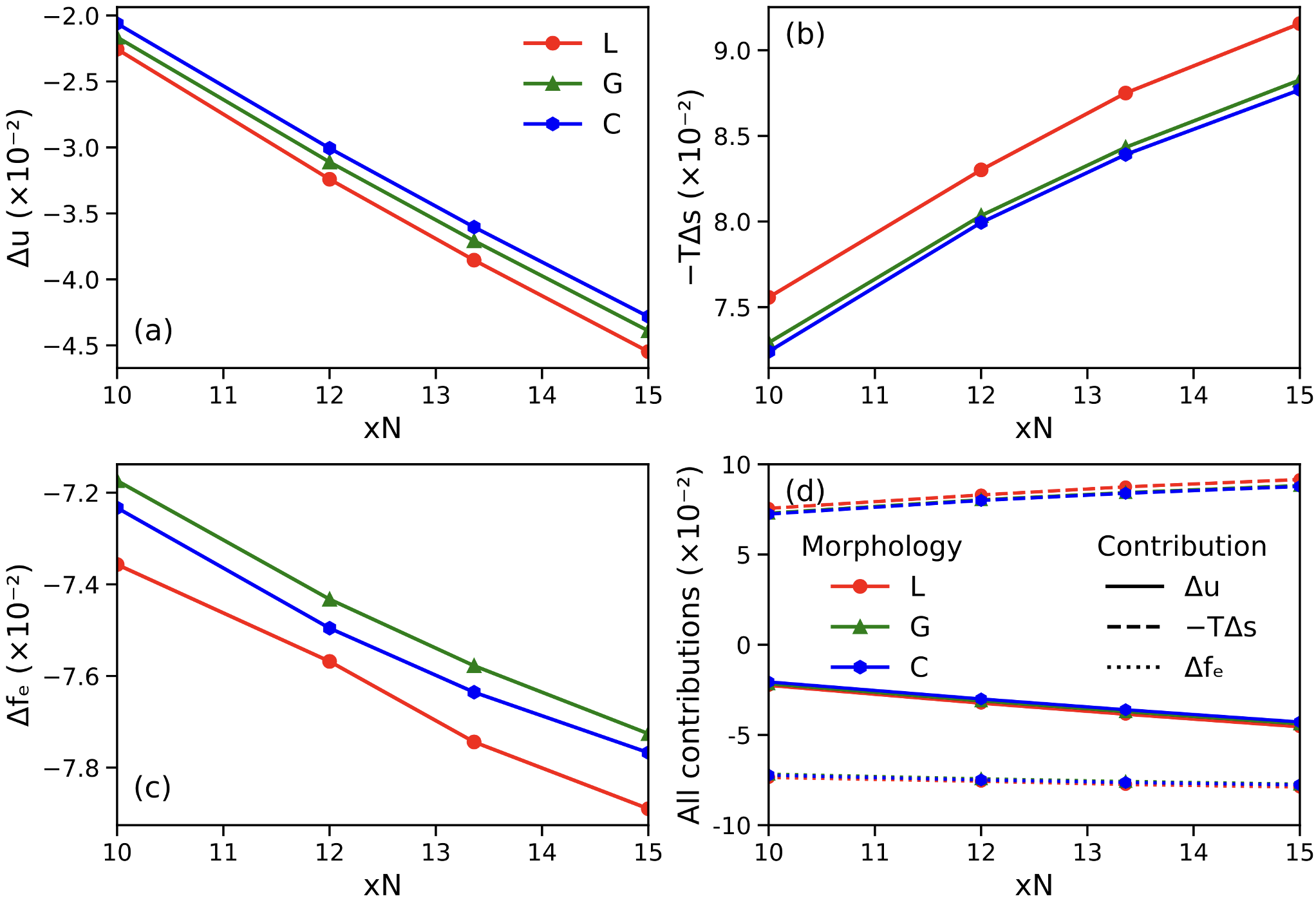}
\caption{Free-energy decomposition for C, G, and L along the C--L phase boundary where $\Delta f({\rm C})=\Delta f({\rm L})$: (a) the mixing enthalpy contribution $\Delta u$, (b) the chain-stretching contribution $-T\Delta s$, (c) the ion-correlation contribution $\Delta f_e$, and (d) all three contributions plotted together. $N=50$.
}
\label{figS4}
\end{figure}

\begin{figure}[H]
\centering
\includegraphics[width=0.48\textwidth]{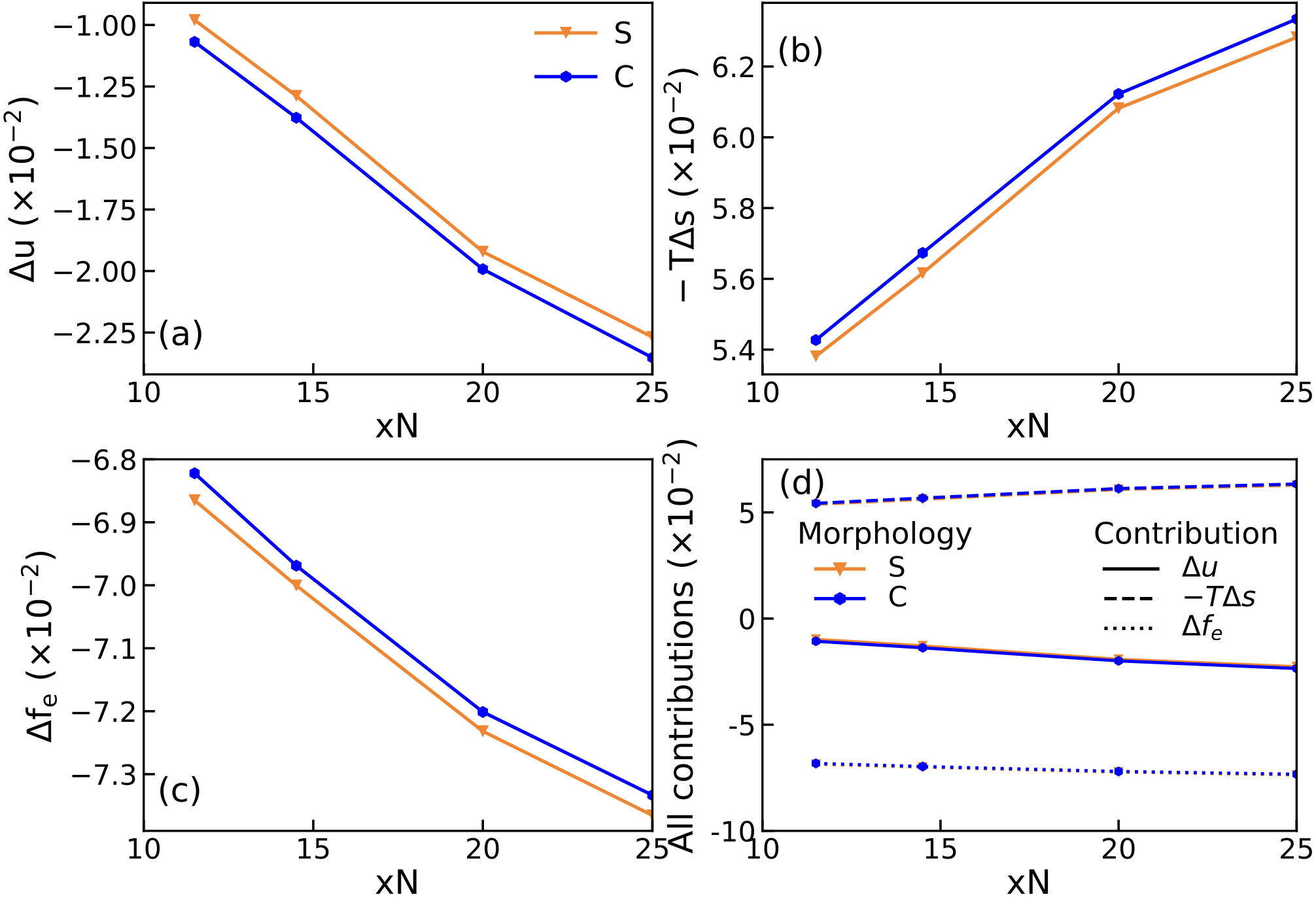}
\caption{ree-energy decomposition for S and C along the S--C phase boundary where $\Delta f({\rm S})=\Delta f({\rm C})$, with each panel defined the same as in Fig.~\ref{figS4}. $N=80$.
}
\label{figS5}
\end{figure}

\begin{figure}[H]
\centering
\includegraphics[width=0.48\textwidth]{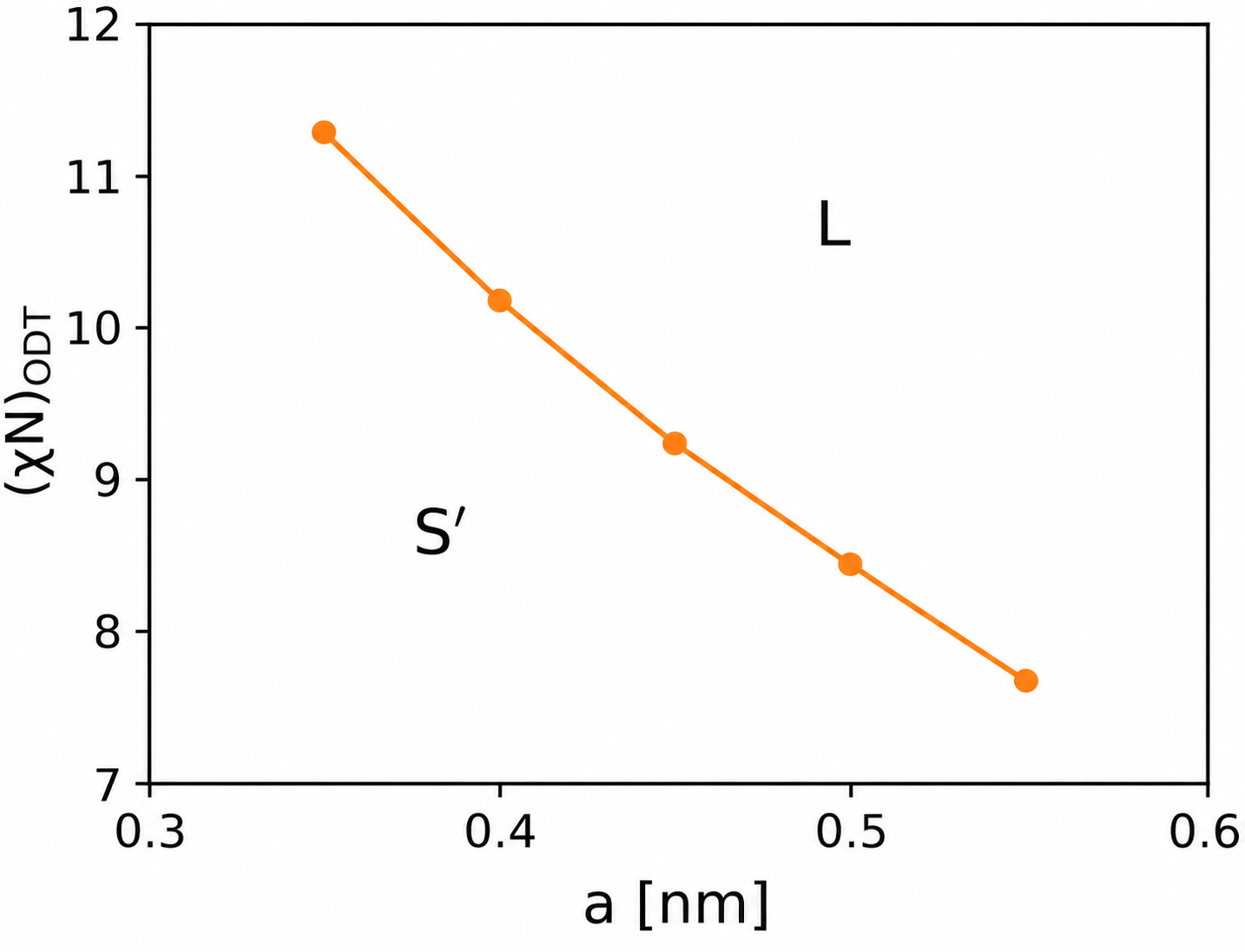}
\caption{Effect of ion size on the order-disorder transition point $(\chi N)_{\rm ODT}$. $N=100$ and $f_A=0.5$.
}
\label{figS6}
\end{figure}

\newpage

\begin{figure*}
\large{For Table of Contents Use Only}
\\
\vspace{1cm}
\large{\bf Microphase Separation and Ionic Aggregation in Oppositely Charged End-Functionalized
Polymer Blends}
\\
\vspace{0.5cm}
Chao Duan and Zhen-Gang Wang
\\
\vspace{2 cm}
\includegraphics[width=0.85\textwidth]{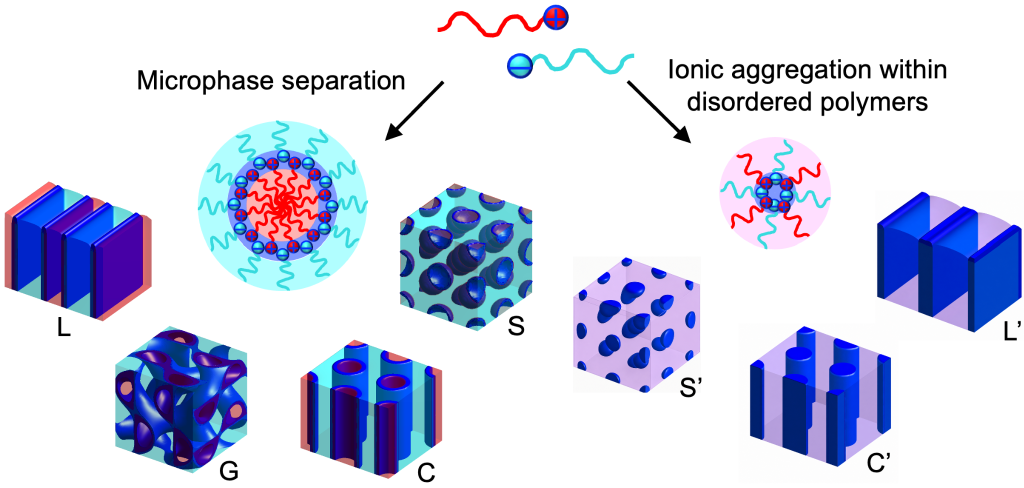}
\end{figure*}

\end{document}